\documentclass{article}
\usepackage{amsmath,url}
 \twocolumn
\makeatletter
\renewenvironment{thebibliography}[1]
     {\small \section*{List of references%
								}%
      \list{\@biblabel{\@arabic\c@enumiv}}%
           {\settowidth\labelwidth{\@biblabel{#1}}%
            \leftmargin\labelwidth
            \advance\leftmargin\labelsep
            \usecounter{enumiv}%
            \let\p@enumiv\@empty
            \renewcommand\theenumiv{\@arabic\c@enumiv}}%
      \sloppy
      \clubpenalty4000
      \@clubpenalty \clubpenalty
      \widowpenalty4000%
      \sfcode`\.\@m}
     {\def\@noitemerr
       {\@latex@warning{Empty `thebibliography' environment}}%
      \endlist\par}
\makeatother

\makeatletter
\long\def\proofbox#1{\gdef\@proofbox{#1}}
\proofbox{\small{\tt\@ifundefined{sourcepath}{}{\sourcepath/}\jobname}\\%
\ifx\UNDEF\Version\else\quad v.\Version, \fi\today}

\def\proofref#1{\proofbox{\small{\tt#1\par
	[edited by Tom Toffoli for personal use]\par
	{\tt\sourcepath/\jobname}, 
	\number\month/\number\day/\number\year
	\par}}}

 \def\affil#1{\\{\small\sl#1\par}}
 \long\def\author#1{\gdef\@author{#1}}
 \author{Tommaso Toffoli ({\tt tt\char"40bu.edu})\affil{Electrical and
Computer Engineering, Boston University, Boston, MA 02215}}

 \long\def\abstract#1{\gdef\@abstract{#1}}
 \abstract{}

\long\def\@firstoftwo#1#2{#1}
\long\def\@secondoftwo#1#2{#2}
\def\@ifundefined#1{%
  \expandafter\ifx\csname#1\endcsname\relax
    \expandafter\@firstoftwo
  \else
    \expandafter\@secondoftwo
  \fi}
 \@ifundefined{leftfrac}{\def\leftfrac{.32}}{}
 \@ifundefined{ritefrac}{\def\ritefrac{.68}}{}

\def\@maketitle{\newpage\noindent\leavevmode
  \begin{minipage}[t]{\leftfrac\textwidth}
    \hrule height0pt
    \@proofbox
  \end{minipage}\hfil
 \begin{minipage}[t]{\ritefrac\textwidth}
    \hrule height0pt
    \raggedleft
    \LARGE\@title\par
    \vskip4pt
    \large\@author
  \end{minipage}
  \vskip8pt
  \ifx\@abstract\@empty\else{\vskip.5em\leftskip1.25in\parskip4pt\small\@abstract\par\vskip.5em}\fi
  \noindent
  \rule{\textwidth}{0.4pt}
  \vskip16pt}
\makeatother

 \makeatletter
 \DeclareRobustCommand\em
        {\@nomath\em \ifdim \fontdimen\@ne\font >\z@
                       \upshape \else \slshape \fi}
\def\@begintheorem#1#2{\sl \trivlist \item[\hskip \labelsep{\bf #1\ #2}]}
\def\@opargbegintheorem#1#2#3{\sl \trivlist
     \item[\hskip \labelsep{\bf #1\ #2\ (#3)}]}
 \newcommand{\sectlabel}[1]{\label{sect:#1}}
 \newcommand{\Sect}[2][]{\def\t@mp{#1}%
\section{#2} \ifx\t@mp\@empty\else\sectlabel{#1}\fi}
 \makeatother

 \newtheorem{theorem}{Theorem}
 \def\qed{\vrule height1.2ex width1ex depth.1ex}
 \newenvironment{tbmatrix}{\bigl[\begin{smallmatrix}}{\end{smallmatrix}\bigr]}
 \newcommand{\twomatrix}[4]{\begin{bmatrix}#1&#2\\#3&#4\end{bmatrix}}
 \newcommand{\ttwomatrix}[4]{\begin{tbmatrix}#1&#2\\#3&#4\end{tbmatrix}}
 \let\epsilon\varepsilon
 \let\rho\varrho
 \let\phi\varphi

 \makeatletter
 \newcommand{\ie}{i.e.,}
 
 \newcommand{\eg}{e.g.,}

 \mathchardef\BY="0202

 \newcommand{\eqlabel}[1]{\label{eq:#1}}
 \newcommand{\Eq}[2][]{\def\t@mp{#1}%
\begin{equation}#2\ifx\t@mp\@empty\notag\else\eqlabel{#1}\fi\end{equation}}
 \newcommand{\Eqaligned}[2][]{\def\t@mp{#1}%
\begin{equation}\begin{aligned}#2\end{aligned}
\ifx\t@mp\@empty\notag\else\eqlabel{#1}\fi
\end{equation}}
 \newcommand{\Eqmultline}[2][]{\def\t@mp{#1}%
\begin{multline}#2\ifx\t@mp\@empty\notag\else\eqlabel{#1}\fi
\end{multline}}
 \newcommand{\Eqgathered}[2][]{\def\t@mp{#1}%
\begin{equation}\begin{gathered}#2\end{gathered}
\ifx\t@mp\@empty\notag\else\eqlabel{#1}\fi
\end{equation}}

 \newcommand{\eq}[1]{(\ref{eq:#1})}	
\makeatother

\def\Tr{\mathrm{Tr}}
\def\Hbar{\overline H}

 \makeatletter
 \renewcommand{\Eq}[2][]{\def\t@mp{#1}%
\begin{equation}#2\ifx\t@mp\@empty\notag\else\eqlabel{#1}\fi\end{equation}}
 \makeatother

 \title{Thermodynamic cost of reversible computing}
 \author{Lev B.\ Levitin and Tommaso Toffoli}

 \abstract{Since reversible computing requires preservation of all
information throughout the entire computational process, this implies that
all errors that appear as a result of the interaction of the
information-carrying system with uncontrolled degrees of freedom must be
corrected. But this can only be done at the expense of an increase in the
entropy of the environment corresponding to the dissipation, in the form of
heat, of the ``noisy'' part of the system's energy.

This paper gives an expression of that energy in terms of the effective noise
temperature, and analyzes the relationship between the energy dissipation
rate and the rate of computation. Finally, a generalized Clausius
principle based on the concept of effective temperature is presented.}

\begin{document}
 \maketitle

\Sect{Cost of computing in the presence of noise}

The concept of reversible computing was introduced in
\cite{bennett:reverse,toffoli:reverse,fredkin:conserve} with the idea to get
rid of the immense energy dissipation and heat generation caused by the
irreversibility of conventional computing processes.  In general, logical
reversibility in computation is a necessary but not sufficient condition for
physical reversibility. However, in quantum computing these two
characteristics come together: any violation of physical reversibility of the
evolution of the controlled information-carrying degrees of freedom
represents noise that, if left unchecked, will destroy the
computation. Thus, it is natural to consider the problem of reversible
computing in a quantum milieu.

Reversible computing is by its nature closer to \emph{communication} than to
conventional computing, and therefore it calls for being analyzed from the
information-theoretical standpoint. Indeed, if it is true that
``communication is computation of the identity function,'' then reversible
computation is \emph{computation of a bijective function}: there is a one-to-one
correspondence between the set of all possible input data and that of all
possible results of the specified computation. Paradoxically, not only does
this requirement make the physical implementation of reversible computation
more difficult, but it also creates a specific cause of energy dissipation
and increase of the entropy of the environment.

Indeed, while at every stage of a conventional computing process we are only
concerned with the integrity of that part of the information that is going to
contribute to the final result (usually, a small fraction of the information
presented in the input data), the precondition of reversible computing is
preservation of \emph{all} information included in the choice of the initial
conditions for any particular instance of computation throughout the the
entire process of computing. However, this goal cannot be achieved for
free. The state of the information-carrying system (the controlled degrees of
freedom) is subject to noise owing to two factors: the statistical nature of
the interaction between the system and the external devices implementing the
required transformation of the system state, and the interactions with
uncontrolled degrees of freedom (the environment). As a result, errors appear
in the state of the system, which requires corrections that inevitably lead
to energy dissipation, \ie\ to an increase of the entropy of the environment.

Henceforth we call `system' the collection of controlled degrees of
freedom involved in the process of computation, rather than the entire
physical object.

\medskip

Suppose that we start with an ensemble $\{\rho_i^0, p_i\}$ of orthogonal,
pure initial states, described by density matrices $\rho_i^0$ that occur with
probabilities $p_i$. Then the initial information $I_0$ (\ie\ the information
in the ensemble of the chosen states $\rho_i^0$ about the ensemble of the
initial data labeled by $i$) is given by
 \Eq{
	I_0 = -\sum_i p_i\ln p_i = -\Tr \rho^0\ln\rho^0,
 }
 where $\rho^0=\sum_i p_i\rho_i^0$ is the a priori density matrix of the
ensemble. The entropy of each initial state is zero. After performing a step
of the computational process (such as, \eg\ application of a ``quantum
gate''), we obtain an ensemble $\{\rho_i, p_i\}$, where the density matrices
$\rho_i$ are in general neither pure nor orthogonal. Note that, in general,
matrices $\rho_i$ may describe states of a different physical system than the
states characterized by $\rho_i^0$. What is important, however, is that there
exists a one-to-one correspondence between the sets $\{\rho_i^0\}$ and
$\{\rho_i\}$.

The average entropy $\Hbar$ of the state after the computation step is
 \Eq{
	\Hbar = - \sum_i p_i\Tr\rho_i\ln\rho_i > 0.
 }
 
Now the information $I$ conveyed by the state ensemble about the initial
state is
 \Eq[ineq_ineq]{
	I \leq H-\Hbar \leq I_0,
 }
  where 
 \Eq{
	H = -\Tr\rho\ln\rho,\quad \text{with}\quad \rho=\sum_ip_i\rho_i.
 }
 Inequality \eq{ineq_ineq} follows from the entropy defect
principle\cite{levitin96}.

In order to achieve reversibility, the initial ensemble of states should
possess sufficient redundancy to turn inequality \eq{ineq_ineq} into an
equality, or, in other words, to make $I= I_0$. Then the output state
$\rho_i$ uniquely determines the corresponding input state $\rho_i^0$. The
reversibility condition allows us to consider a ``reverse channel'' with
ensembles $\{\rho_i,p_i\}$ as input and $\{\rho_i^0,p_i\}$ as output. In this
channel, the quantity $\Hbar$ plays the role of
\emph{equivocation}\cite{shannon48}.  It expresses the effect of noise and
the presence of errors in the results of the computation. According to
Shannon's $10^\text{th}$ theorem (the ``correction channel
theorem''\cite{shannon48}), $\Hbar$ is equal to the minimum additional
information required in order to correct all the errors in the state of the
system and thereby preserve the initial amount of information (though, in
general, in a transformed form). In other words, to correct the errors, the
entropy of the information-carrying system must be decreased by the value of
$\Hbar$. Of course, this can be done only at the price of an entropy increase
of at least $\Hbar$ in other degrees of freedom (\eg\ the environment). The
actual amount of energy that has been dissipated within the system depends on
the concrete properties of the system itself, as shown by the following
considerations (cf.\ \cite{levitin98}):

Denote by $E(T')$ the energy of the system in the state of thermal
equilibrium at temperature $T'$. The increase of the average entropy $\Hbar$
means that part of the energy has been transformed into heat in the system
or transferred to the system in the form of heat. How large is this part? It
is equal to the energy $E(T)$ that the system would have had in the state of
thermal equilibrium at temperature $T$, where $T$ is determined by the value
$\Hbar$ of entropy increase according to the relation
 \Eq[ent_inc]{
	\Hbar = H(T) = \int_0^T \frac1{kT'}\frac{dE(T')}{dT'} dT';
 }
 thus, $T$ has the meaning of an \emph{effective noise temperature}.

The concept of effective noise temperature was first introduced in
\cite{levitin98}. Being applicable to any non-equilibrium state of a system,
the effective temperature has, nevertheless, the same fundamental properties
as the usual temperature in the thermodynamics of equilibrium systems (and
coincides with latter at equilibrium).  In particular, it can be shown that if
two non-equilibrium physical systems have effective temperatures $T_1$ and
$T_2$, where $T_1<T_2$, and corresponding ``thermal'' energies $E_1(T_1)$ and
$E_2(T_2)$ as defined by \eq{ent_inc}, it is impossible to transfer an amount
of energy $\Delta E$ from the first system to the second by decreasing
$E_1(T_1)$ by an amount $\Delta E$ and correspondingly increasing $E_2(T_2)$
by the same or a smaller amount as a sole result of a physical process. In
fact, this property represents a \emph{generalized Clausius
principle}\cite{clausius}. One is thus justified in regarding $E(T)$ as an
amount of energy converted into heat.

In the absence of noise ($\Hbar=0$), after one step the system will find
itself in the pure state corresponding to the initial state $\rho_i^0$. If
$\Hbar\neq0$, this implies a nonzero probability of being in a state
different from the correct one---the probability of error. Correspondingly,
in the calculation of $E(T)$, if $\Hbar=0$, then $T=0$ and $E(T)=0$, \emph{as
if} the system were in its ground state. If $\Hbar\neq0$, then $T\neq0$ and
$E(T)\neq0$.  $E(T)$ can be expressed alternatively as a function of the
entropy $\Hbar$, or of the probability of being in a non-ground state---which
we interpret below as the probability of error.

It follows from the above considerations that the procedure of error
correction entails removing an amount of heat $E(T)$ from the system and
applying to it an amount of work required to restore the correct state.

Even though it arises from disturbances introduced into a system by
interactions with its environment, the noise temperature is not equal, in
general, to the temperature of the environment itself (were the latter at
thermal equilibrium).

\medskip

Note that the minimum amount of heat $Q_\text{e}$ to be ultimately transfered
at every step to an environment at temperature $T_\text{e}$ (an ``infinite,
constant-temperature heat sink'') to ensure continuous, closed-cycle
operation of the invertible computing machinery envisaged here is not equal,
in general, to $E(T)$. In fact, in a reversible process, $Q_\text{e}=\Hbar
T_\text{e}$. Indeed, if $T_\text{e}<E(T)/\Hbar$ we could even produce some
``useful work'' by removing heat $E(T)$ from our system. However, in a real
situation, usually $T_\text{e}\geq E(T)/\Hbar$ (we cannot allow the computing
system to ``heat up'' too much since too large a probability of error would
make the state of the system incorrectible) and thus $Q_\text{e}\geq E(T)$.

\medskip

One could reason that the amount of dissipated energy can be made arbitrarily
small if one works with ``low energy'' states. But it should be borne in
mind that the lower the energy, the larger the time taken by each computational
step. As shown in \cite{margolus98,levitin03,levitin02}, the minimum time for
transforming a state to an orthogonal state (``flipping a qubit'') is
 \Eq{
	\tau\geq \frac h{4E},
 }
 where $h$ is Plank's constant and $E$ the average energy of the
system. This inequality turns into an equality only for a system with two
orthogonal states. For a sequence of $N$ mutually orthogonal states a
stronger inequality,
 \Eq{
	\tau \geq \frac{N-1}N\frac h{2E},
 }
 becomes valid\cite{margolus98}. Thus, the maximum rate of computation, \ie\ the
number $R$ of computational steps per unit time is proportional to the
quantum-mechanical average energy of the system, that is,
 \Eq[two]{
		R = \frac{4E}h
 }
 for two orthogonal states and
 \Eq[many]{
		R = \frac N{N-1}\frac{2E}h
 }
 for a sequence of $N$ such states.

As shown below, it follows from \eq{ent_inc}, \eq{two}, and \eq{many} that,
the faster one wants to perform a computation, the more energy \emph{per
step} will be dissipated in the form of heat.  The energy dissipation
per computational step is then expressed as an increasing function of both the
noise temperature and the rate of computation (\ie\ the number of
computational steps per unit time).

\Sect[examples]{Examples}

We shall consider first two representative examples.

\medskip

\noindent{\sc Example 1}. The qubit.

\smallskip

Let a system with two orthogonal states (a two-dimensional Hilbert space)
have two energy levels, $E_0=0$ and $E_1$. The maximum rate of computation is
achieved for two pure states $\psi_1=\frac1{\sqrt2}(|E_0\rangle+|E_1\rangle)$
and $\psi_2=\frac1{\sqrt2}(|E_0\rangle-|E_1\rangle)$, with quantum-mechanical
average energy $E = E_1/2$ (cf.\ \cite{margolus98}), which
turn into one another at each computational step. Their density matrices in the
stationary basis $\{|E_0\rangle,|E_1\rangle\}$ are
 \Eq{
  \rho_1^0 = \twomatrix{1/2}{1/2}{1/2}{1/2}\quad\text{and}\quad
  \rho_2^0 = \twomatrix{1/2}{-1/2}{-1/2}{1/2}.
 }
 In this case
 \Eq[nu_avg]{
	R = \frac{4E}h = \frac{2E_1}h.
 }

Suppose now that there is a probability $\epsilon$ of error, so that the
resulting states are each a mixture of the correct state with probability
$1-\epsilon$ and the other state with probability $\epsilon$, with density
matrices
 \Eq[spread]{
  \rho_1 = \twomatrix{1/2}{1/2-\epsilon}{1/2-\epsilon}{1/2},\ 
  \rho_2 = \twomatrix{1/2}{-1/2+\epsilon}{-1/2+\epsilon}{1/2}.
 }
 The average entropy of the states is 
 \Eq{
	\Hbar = -\epsilon\ln\epsilon-(1-\epsilon)\ln(1-\epsilon),
 }
 and is equal to the entropy of a thermal equilibrium state with
temperature $T$ described by the density matrix
 \Eq[ratio]{
	\rho_\text{eq}=\twomatrix{1-\epsilon}00\epsilon,\quad\text{where}\quad
 \epsilon = \frac{e^{-E_1/kT}}{1+e^{-E_1/kT}}.
 }

 The states in \eq{spread} and \eq{ratio} are thermodynamically equivalent,
since they have the same entropy $\Hbar$. In particular, they can be
represented by mixtures of a pure state and the maximum-entropy state
 $\ttwomatrix{1/2}00{1/2}$
 with the same coefficients, respectively, $1-2\epsilon$ and $2\epsilon$:
 \Eqaligned{
 \rho_i &= (1-2\epsilon)\twomatrix{1/2}{\pm1/2}{\pm1/2}{1/2}+
	2\epsilon\twomatrix{1/2}00{1/2},\quad i=1,2;\\
 \rho_\text{eq} &= (1-2\epsilon)\twomatrix1000+
	2\epsilon\twomatrix{1/2}00{1/2}.
 }
 Thus, $T$ in \eq{ratio} is the effective noise temperature as defined by
\eq{ent_inc}.

The energy of the system at the thermal equilibrium state with temperature
$T$ is
 \Eq[erg_eq]{
	E(T) = \frac{e^{-E_1/kT}}{1+e^{-E_1/kT}}E_1.
 }
 Hence, from \eq{nu_avg} and \eq{erg_eq},
 \Eq{
	E(T) = \frac{e^{-hR/2kT}}{1+e^{-hR/2kT}}\frac{hR}2,
 }
 and thus, by \eq{ratio},
 \Eq{
	E(T) = \frac{hR}2\epsilon.
 }

The quantity $E(T)$ represents the lower bound on energy dissipation per step
of computation---in other words, per ``flipping of the qubit.'' Thus, for a
qubit, assuming a fixed probability of error, the minimum energy dissipation
is proportional to the rate of computation. Consequently, the amount of heat
$Q$ generated per unit of time is proportional to the square of the rate of
computation:
 \Eq[res1]{
	Q = \frac{hR^2}2\epsilon
 }

\medskip

\noindent{\sc Example 2}. Quantum harmonic oscillator.

\smallskip

A sequence of $N$ orthogonal non-stationary states of a quantum harmonic
oscillator consists of wavefunctions
 \Eq[levels]{
   \psi_m {=}\!\! \sum_{n=0}^{N-1}\frac1{\sqrt{N}} e^{-2i\pi mn/N} |E_n\rangle,
 \quad m,n{=}0,1,\dots,N{-}1,
 }
 where $|E_n\rangle$ is the stationary state with energy $E_n=n\Delta E$ and
$\Delta E$ is the separation of the energy levels (the ground-state energy is
taken to be zero). The average energy is \hbox{$E=(N-1)\Delta E/2$}.

Sequence \eq{levels} provides the maximum rate of computation, given by
 \Eq{
	R = \frac{2N}{N-1}\frac Eh = N\frac{\Delta E}h.
 }

Assume now that, as a result of errors caused by noise, the resulting state is
not pure, but has an entropy $\Hbar$. To calculate the part of the energy
turned into heat, consider the state of thermal equilibrium with the same
entropy. The energy corresponding to the thermal equilibrium of a harmonic
oscillator at temperature $T$ is
 \Eq{
	E(T) = \frac{\Delta E}{e^{\Delta E/kT} - 1}.
 }
 Hence, for a given maximum computation rate, the energy dissipation per
computational step is
 \Eq{
	E(T) = \frac{hR/N}{e^{hR/NkT}-1}.
 }
 The error probability (the total probability to be in a state which is orthogonal to the correct one) is in this case
 \Eq{
  \epsilon = e^{-\Delta E/kT} = e^{-hR/NkT}.
 } 
 Hence, in terms of error probability,
 \Eq{
	E(T) = \frac\epsilon{1-\epsilon}\frac{hR}N.
 } 
 Thus, the rate of energy dissipation (\ie\ heat production per unit time) is
 \Eq[res2]{
	Q = \frac\epsilon{1-\epsilon}\frac{hR^2}N.
 }

\bigskip

Through expressions \eq{res1} and \eq{res2}, the above two examples suggest a
general conjecture---that for a fixed error probability (\ie\ for a given
intensity of noise effects) the energy dissipation rate increases
\emph{quadratically} with the rate of computation.

In the next section we shall confirm this conjecture in the limiting
case of a large number of degrees of freedom.

\Sect{General case: a system with many degrees of freedom}

The exact form of the expression for heat production depends on the specific
details of the computing system's energy-level structure. Nevertheless, under
rather general assumptions, a closed-form result can be obtained for a broad
class of systems.

Following the analysis given in \cite{margolus98}, let's consider a system
with many degrees of freedom that runs through a long sequence ($N\gg1$) of
mutually orthogonal states. It is shown in \cite[Sect.\ 2.3]{margolus98} that
all those states are superpositions of energy eigenstates with all different
values of energy,
 \Eq{
	|\psi_m\rangle = \sum_{n=0}^{N-1}c_ne^{-2\pi i E_n/E_N} |E_n\rangle,
							\quad m=0,1,\dots,N-1,
 }
 where
 \Eq{
	c_n = \sqrt{\frac{E_{n+1}-E_n}{E_N}}
 }
 and the average energy $E$ is asymptotically ($N\gg1$)
 \Eq[Easympt]{
	E = \langle\psi_m|H|\psi_m\rangle = E_N/2.
 }

Let $T$ be the effective noise temperature of our system. The partition
function $Z$ and the average energy $E$ of the thermal equilibrium state are
 \Eq[part]{
	Z=\sum_{n=0}^\infty w(E_n)e^{-E_n/kT},
 }
 \Eq[Epart]{
	E(T)=\frac1Z \sum_{n=0}^\infty E_n w(E_n)e^{-E_n/kT},
 }
 where $w(E_n)$ is the number of the microstates with energy $E_n$.  For a
system with many degrees of freedom one has $w(E_n)\sim E_n^\alpha$ (for
$n>0$), where $\alpha\geq1$. Typically\cite{reif65}, $\alpha\gg1$. Also, we
assume that the zero-energy level is non-degenerate, \ie\ $w(E_0)=1$.

\medskip

\def\c{a}	

Let us introduce a dimensionless variable $x=NE/E_N$ that characterizes the
density of the energy levels: the number of energy levels between $x_n$ and
$x_n+\Delta x$ is equal to $\Delta x$. Also, suppose that $w(E_n)=\c
x_n^\alpha$, where $\c$ is a dimensionless constant. Then expressions
\eq{part} and \eq{Epart} take the following form:
 \Eq[part']{
 	Z=1+\sum_{n=0}^\infty \c x_n^\alpha e^{-(E_N/NkT)x_n},
 }
 \Eq[Epart']{
 E(T)= \frac1Z \sum_{n=0}^\infty \c x_n^{\alpha+1}\frac{E_n}N e^{-(E_N/NkT)x_n}.
 }
 It follows from \eq{many} and \eq{Easympt} that, for $N\gg1$, one has
$E_N=hR$. Replacing summation by integration in \eq{part'} and \eq{Epart'},
we obtain, using Euler's gamma function $\Gamma$,
 \Eq{\begin{aligned}
	Z &= 1+\int_0^\infty \c x^\alpha e^{-(hR/NkT)x} dx \\
	  &= 1+\c\Gamma(\alpha+1)\left(\frac{NkT}{hR}\right)^{\alpha+1},
     \end{aligned}
 }
 \Eq{\begin{aligned}
  E(T) &= \frac1Z \int_0^\infty \c x^{\alpha+1}\frac{hR}N e^{-(hR/NkT)x} dx \\
       &= \frac1Z\, \c\Gamma(\alpha+2)kT \left(\frac{NkT}{hR}\right)^{\alpha+1}.
     \end{aligned}
 }

 It will be convenient to express this result in terms of the error
probability $\epsilon$, which is equal to the probability to be in a
non-ground state, that is,
 \Eq[error]{
	\epsilon = 1-\frac1Z;
 }
 then
 \Eq{
	E(T)=(\alpha+1)kT(1-\frac1Z)=(\alpha+1)kT\epsilon.
 }
 Expressing $kT$ in terms of $\epsilon$ from \eq{error}, we find that
 \Eq[E(T)]{
 E(T)=\frac{(\alpha+1)\epsilon}N
\left[\frac\epsilon{1-\epsilon}\cdot\frac1{\c\Gamma(\alpha+1)}\right]^{\frac1{\alpha+1}} hR\,.
 }
 Thus, the energy dissipation per computational step is proportional to the
rate of computation, and so
 \Eq[Q]{
	Q\sim hR^2,
 }
 \ie\ the \emph{rate of heat production} is proportional to the \emph{square
of the computation rate}.

\section*{Appendix: A generalized Clausius principle}

\begin{theorem} Given two (in general, non-equilibrium) physical systems with
entropies $H_1$ and $H_2$, thermal equilibrium functions $E_1(T)$ and $E_2(T)$,
and effective temperatures $T_1$ and $T_2$, defined by equations
 \Eq{
 H_i = \int_0^{T_i}\frac1{kT} \frac{dE_i(T)}{dt} dt,\quad(i=1,2)
 }
 it is impossible to transfer energy from the first system to the second so
as to decrease $E_1(T_1)$ by an amount of energy $\Delta E$ and increase
$E_2(T_2)$ by the same or a smaller amount as a sole result of a physical
process.  We assume that the functions $E_1, E_2$ that characterize the
systems remain unchanged as a result of the process.
 \end{theorem}

\noindent{\bf Proof.}\quad Suppose $E_2(T_2)$ has been decreased by $\Delta
E$ and $E_2(T_2)$ increased by $\Delta E'\leq\Delta E$ without any other
changes in the environment. As a result, the systems would now have new
effective temperatures $T_1'<T_1$ and $T_2'>T_2$, and corresponding thermal
equilibrium energies
 \Eq{\begin{aligned}
	E_1(T_1') &= E_1(T_1)-\Delta E,\\
	E_2(T_2') &= E_2(T_2)+\Delta E',
     \end{aligned}
 }
where
 \Eq{
	 \Delta E'=\int_{T_2}^{T_2'}\ \frac{dE_2(T)}{dT}\,dT
     \leq\Delta E =\int_{T_1'}^{T_1} \frac{dE_1(T)}{dT}\,dT
 }
 (we assume that $E_1$ and $E_2$ are monotonically increasing functions).
Hence, the total change of entropy would be
 \Eq{
	\Delta H = \Delta H_1+\Delta H_2,
 }
 where
 \Eq{
  \Delta H_1=-\int_{T_1'}^{T_1}\frac{dE_1(T)}{TdT}\,dT,
 }
 \Eq{
  \Delta H_2 = \int_{T_2}^{T_2'}\frac{dE_2(T)}{TdT}\,dT.
 }
  Using the mean value theorem, we get
 \Eq[A1]{
	\Delta H_1 = -\frac1{T_1^*}\Delta E,\quad T_1'<T_1^*<T_1,
 }
 \Eq[A2]{
	\Delta H_2 = \frac1{T_2^*}\Delta E',\quad T_2<T_2^*<T_2'.
 }
 Obviously, $T_1^*<T_1\leq T_2<T_2^*$. 

It follows from \eq{A1}, \eq{A2} that the total change of entropy would be
negative:
 \Eq{\Delta H = -\frac1{T_1^*}\Delta E + \frac1{T_2^*}\Delta E'
	\leq\frac{T_1^*-T_2^*}{T_1^*T_2^*} \Delta E \leq 0.
 }
 Thus, such a process is impossible.\quad\qed

 \medskip

Note that the generalized Clausius principle does not preclude any exchange
of work between the two systems, \ie\ a process that does not affect the
values of $E_1(T_1)$ and $E_2(T_2)$.

The significance of these results is that they represent a \emph{universal}
lower bound on the energy dissipation rate in a reversible computation
process.

\section*{Acknowledgment}

The authors are very grateful to Michael Frank ({\sc famu--fsu}) whose deep and
thoughtful comments resulted in improvements of this paper and gave us
inspiration for future work.


\begin{thebibliography}{99}
\renewcommand{\itemsep}{0pt}
 \let\bib\bibitem
 \bib{bennett:reverse} {\sc Bennett}, Charles, ``Logical reversibility of
computation,'' {\sl IBM J.\ Res. Develop. \bf6} (1973), 525--532.
 \bib{clausius}{\sc Clausius}, R., {\sl The Mechanical Theory of Heat},
MacMillan 1879.
  \bib{fredkin:conserve}{\sc Fredkin}, Edward, and Tommaso {\sc Toffoli},
``Conservative logic,'' {\sl Int. J. Theoret. Phys. \bf21} (1982), 219--253,
 \bib{levitin96}{\sc Levitin}, Lev B., ``On the quantum measure of
information,''in {\sl Proc. 4th Conf. on Information Theory}, Tashkent 1969,
111--116. English translation in {\sl Annales de la Foundation Louis de
Broglie \bf21}:3, (1996), 345--348.
 \bib{levitin98}{\sc Levitin}, Lev B., ``Energy cost of information
transmission (along the path to understanding),'' {\sl Physica D \bf120}
(1998), 162--167.
 \bib{levitin02}{\sc Levitin}, Lev B., Tommaso {\sc Toffoli}, and Zachary
{\sc Walton}, ``Maximum speed of quantum gate operation,''
{\sl Int. J. Theoret. Phys. \bf44} (2005), 965--970.
 \bib{levitin03}{\sc Levitin}, Lev B., Tommaso {\sc Toffoli}, and Zachary
{\sc Walton}, ``Operation time of quantum gates,'' {\sl Quantum
Communication, Measurement, and Computing} (J. H. {\sc Schapiro} and O. {\sc
Hirota}, ed.), Rinton 2003, 457--459.
 \bib{margolus98}{\sc Margolus}, Norman, and Lev B. {\sc Levitin}, ``Maximum
speed of dynamical evolution,'' {\sl Physica D \bf120} (1998), 188--195.
 \bib{reif65}{\sc Reif}, Federick, {\sl Fundamental of Statistical Mechanics},
McGraw-Hill 1965.
 \bib{shannon48}{\sc Shannon}, Claude, ``A mathematical theory of
communication,'' {\sl Bell System Technical J. \bf27} (1948), 379--423 and
623--656.
 \bib{toffoli:reverse} {\sc Toffoli}, Tommaso, ``Reversible computing,'' {\sl
Seventh Colloquium on Automata, Languages and Programming}, J.~W.~{\sc de
Bakker} and J.~{\sc van~Leeuwen}, eds., Springer 1980, 632-644.

\end{thebibliography}
\end{document}